\documentstyle[11pt]{article}

\makeatletter
\renewcommand{\@cite}[2]{}

\renewcommand{\@biblabel}[1]{#1.\hfill}

\input epsf

\begin{document}
\large

Astronomy Reports, Vol. 46, No. 8, 2002, pp. 639 645. Translated
from Astronomicheski.. Zhurnal, Vol. 79, No. 8, 2002, pp. 708-714.
Original Russian Text Copyright $C_{2002}$ by Shatskiy, Kardashev.
$$ $$
\begin{center}
{\bf An Induction Accelerator of Cosmic Rays\\
on the Axis of an Accretion Disk.
$$ $$
A. A. Shatskiy   and   N. S. Kardashev}

Astro Space Center, Lebedev Physical Institute, Moscow, Russia
Received January 3, 2002; revised February 1, 2002.
$$ $$
{\bf\qquad ABSTRACT}
\\
The structure and magnitude of the electric field created by a
rotating accretion disk with a
poloidal magnetic field is found for the case of a vacuum
approximation along the axis. The accretion disk is
modeled as a torus filled with plasma and the frozen-in magnetic
field. The dimensions and location of the
maximum electric field are found, as well as the energy of the
accelerated particles. The gravitational field is
assumed to be weak.
\end{center}

\section{INTRODUCTION}

Recently, there has been wide discussion of various
mechanisms for accelerating particles around
supermassive black holes (SMBHs) in the nuclei of
galaxies and stellar-mass black holes in our Galaxy,
in connection with studies of synchrotron radiation
and inverse Compton scattering in the well collimated
jets observed from radio to gamma-ray wavelengths.
Extremely high-angular-resolution observations obtained
via radio interferometry show that these jets
become very narrow (comparable to the gravitational
radius) with approach to the black hole. Explanations
of particle acceleration near relativistic objects
(black holes and neutron stars) are usually based
on two types of mechanisms: acceleration by electric
fields and magnetohydrodynamical acceleration (the
Blandford-Znajek mechanism [1]). Acceleration by
an electric field, and the very existence of the electric
field, are inseparably linked with the low density
of plasma in this volume. Conditions justifying
a vacuum approximation are probably realized in
the magnetospheres of pulsars and in some types of
SMBHs [2, 3]. In this case, it is possible to accelerate
particles to extremely high energies [4]. The limiting
charge densities for which the vacuum approximation
remains valid are determined by the formula [5]
\begin{equation}
n_e <  {|({\bf \Omega H})|\over 2\pi c e}\simeq (1\,\, день /P)\cdot
(H/10^4 Гс)\cdot 10^{-2} \, см^{-3}\, .
\label{0-1}\end{equation}
Here, $\Omega$  is the angular velocity of rotation, $H$ the
characteristic magnetic-field strength, $P$ the rotational
period, $c$ the velocity of light, and $e$ the electron
charge. It is evident from this expression that
we should have for typical quasars
$n_e < 10^{-2} cm^{-3}$.
Note that, in intergalactic space,
$n_e\approx 10^{-6} cm^{-3}$,
and in the Galaxy,
$n_e\approx 1 cm^{-3}$.
The presence of
a black hole in the center of the Galaxy also leads
to a decrease in $n_e$ near the center. In addition, the
magnetic fields near SMBHs can reach values of the
order of $10^9 G$ [4]. In any case, the question of the
applicability of the vacuum approximation is rather
complex, and must be solved taking into account the
physics of black holes.
In this paper, we will assume that the conditions
for the vacuum approximation are satisfied; this will
enable us to investigate the structure of the electric
field excited by a rotating accretion disk with a
poloidal magnetic field. The formulation of this problem
is analogous to that considered by Deutsch [6],
who found the structure of the electric field created by
a dipolar magnetic field frozen in a rotating star. If a
conductor rotates together with a frozen-in magnetic
field, then, in a rotating coordinate system in which
the conductor is at rest, there must be no electric field
inside the conductor. Therefore, in an inertial system,
an electric field is induced inside the conductor due
to the presence of the magnetic field, and this electric
field gives rise to a surface charge (in the special case
of a magnetic dipole with a quadrupolar distribution).
This surface charge is the source of the external electric
field. We will consider the analogous problem for
an accretion disk.

\section{CONSTRUCTION OF THE MODEL}

\begin{figure}[t]
\centering \epsfbox[100 500 400 700]{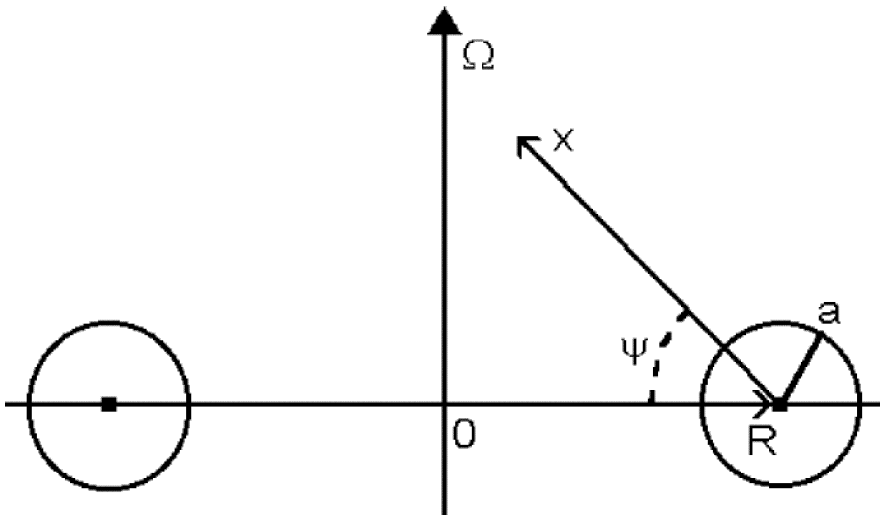} \label{r1}
\caption{Accretion disk in the form of a torus (side view).}
\end{figure}

Let us consider a stationary system consisting
of a rotating accretion disk in the form of a regular
torus filled with plasma and surrounded by a poloidal
magnetic field. We neglect motion of the torus toward
the center. The geometry of the torus is determined
by the two radii $a$ and $R$ (Fig. 1). Due to the high
conductivity of the plasma, the magnetic field is frozen
inside the torus, and currents flow only in the toroidal
direction; there is no matter outside the torus. In
a coordinate system comoving with the plasma, the
electric field vanishes due to the relation
\begin{equation}
{j'}_\alpha =\sigma' {F'_{\alpha 0}} ,
\label{1-1}\end{equation}
where $j'_\alpha = 0$ are the poloidal components of the
current density,
$\sigma'$ is the plasma conductivity, and
$F'_{\alpha 0}$ are the covariant components of the electric field
tensor in a system comoving with the plasma. We
introduce the following coordinates (Fig. 1):
$x$ is the distance from an arbitrary point to the center inside
the torus in the same meridional plane, $\psi$  is the angle
between the direction toward the center of the system
and the direction to a given point from the center
inside the torus in the same meridional plane, and
$\varphi$
is the position angle defining this plane. Thus, the
differential coordinates $dx$, $d\psi$, and $d\varphi$ form a
righthanded orthogonal vector triad. We can write the
square of a linear element in Minkowski space for the
differentials of these coordinates
\footnote{We take the velocity of light to be unity: $c = 1$.}:
\begin{equation}
\left\{
\begin{array}{rcl}
ds^2 =dt^2 -dx^2 - x^2\, d\psi^2 -(R-x\cos\psi )^2\, d\varphi^2\, ,\\
\sqrt{-g}=x |R-x\cos\psi | .\end{array}\right.
\label{1-2}\end{equation}
Let the plasma in the torus rotate with angular velocity
$\Omega$
 with respect to a distant observer. Then, the
coordinate transformation is given by
\footnote{Where not indicated otherwise,
$x_i = t, x, \psi , \varphi$,  аnd $x_\alpha =x, \psi$.}
\begin{equation}
dx^i = dx'^k[\delta^i_k+\Omega\delta^i_\varphi\delta^0_k]
\label{1-3}\end{equation}
We now introduce the covariant four-vector potential
of the electromagnetic (EM) field $A_i$. Due to the axial
symmetry of the system, only $A_0$, the potential of the
electric field, and $A_\varphi$, the potential of the magnetic
field, differ from zero. Therefore, the EM-field tensor
$F_{ij}$ has only poloidal components:
\begin{equation}
F_{\alpha 0}=\partial_\alpha A_0 \, ,
\quad F_{\alpha\varphi}=\partial_\alpha A_\varphi .
\label{1-4}\end{equation}
These components transform in accordance with (4)
[7, $\S$83]:
\begin{equation}
F'_{\alpha 0}=F_{\alpha 0} +\Omega F_{\alpha\varphi}\, ,
\quad F'_{\alpha\varphi}=F_{\alpha\varphi}
\, ,\quad A'_0 = A_0 + \Omega A_\varphi \, ,
\quad A'_\varphi = A_\varphi \, .
\label{1-5}\end{equation}
Since $F'_{\alpha 0} = 0$ in the plasma [see (2)], we have for
$x < a$:
\begin{equation}
F_{\alpha 0}=-\Omega F_{\alpha\varphi}\, .
\label{1-6}\end{equation}
It follows from (5) and (7) that, when
$x < a$, $A_0 = const - \Omega A_\varphi$.
 It follows from the third equation of (6)
that this constant is $A'_0$ inside the torus.

The continuous boundary conditions for the tangential
electric and normal magnetic components of
the EM field act at the interface between the plasma
and vacuum (at the surface of the torus). These components
should vanish at the equator, due to the
axial symmetry of the system and the mirror
(anti-)symmetry of the components of the EM field.
Hence, the normal component of the magnetic field can be
expanded in a Fourier sine series:
\begin{equation}
F_{\psi\varphi}=n R_n(x) \sin (\psi n) .
\label{2-1}\end{equation}
Here and below, the summation over $n$ is assumed,
where $n$ runs through all numbers of the natural
series. It follows from (7) and (8) that the boundary
condition for the tangential electric field is
\begin{equation}
F_{\psi 0}{}_{(a,\psi )}=-\Omega n R_n(a)\sin (\psi n)\, .
\label{2-2}\end{equation}

\section{POTENTIAL AND STRUCTURE
OF THE ELECTRIC FIELD
NEAR THE TORUS}

Near the torus [see formulas (30), (32), and (37) in
the Appendix], the main contribution to the potential
$A_0$ is made by the first term of the first harmonic of
the Fourier series. Far from the torus, the potential
dies away. The kinetic energy of a charged particle
accelerated by the system is determined primarily by
this part of the potential. Accordingly, we obtain the
main approximation for the difference in the potentials
at the torus surface between the angles
$\psi$ and $\psi = \tilde\psi$:
\begin{equation}
\Delta A_0 = \Omega\cdot R\cdot H_0\cdot a\cdot [\ln(4/b) -1]
\cdot [1-\cos\tilde\psi ]/\pi  \, .
\label{5-1}\end{equation}
This same expression can also be obtain in another
way [8, $\S$63]. The corresponding invariant result has
the form
\begin{equation}
\varepsilon\equiv U^i \Delta A_i =
\int\limits_C [{U}^i\, F_{j i}]\, dx^j =
\int\limits_0^{\tilde\psi} {U'}^\varphi F'_{\psi\varphi}\, d\psi \, .
\label{5-1-1}\end{equation}
Here, $U^i$ denotes the four-velocity of the observer
\footnote{Who is at rest relative to the distant stars.}
at the measurement point, and the contour for the
integral is chosen for convenience to be on the inside
surface of the torus in a reference system comoving
with the torus. After substituting into this expression
${U'}^\varphi \approx \Omega$
and the expression for
$F'_{\psi\varphi} = F_{\psi\varphi}$
we obtain (10).


%
\begin{figure}[t]
\centering \epsfbox[10 550 400 700]{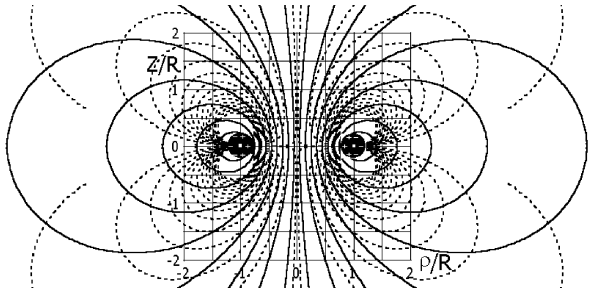} \label{r2}
\caption{Appearance of the electromagnetic field lines in the
system. The solid curves show the magnetic field lines and the
dashed curves the electric field lines. $z$ and $\rho$ are
expressed dashed curves the electric field lines. $z$ and $\rho$
are expressed in fractions of $R$.}
\end{figure}

The electric field inside the torus is zero, since the
torus is conducting. In an inertial reference system,
this is accomplished by the compensation of two
fields: that induced by the rotation of the magnetic
field and the field of the surface charge on the torus.
The surface density of the electric charge on the torus
is  $\rho_e = - F_{x0}/(4\pi)$.
$F_{x0}$ is the normal component of
the electric field on the torus surface [see (7)]. At
large distances from the torus surface ($x >> a$), these
charges represent a superposition of dipoles (Fig. 2).
However, at distances $x >> R$, the field from all these
dipoles has a quadrupole character. It is not difficult to
obtain an expression for the dipole moment per unit
angle $\varphi$:
\begin{equation}
d={\Omega R^2 a^2 H_0\over 2\pi}\left[ 2 - \ln (4/b)\right] .
\label{5-1-2}\end{equation}
The electric field and its potential $\phi$ can be obtained
by integrating all the dipoles over the angle $\varphi$ [7, 40].
As a result, we obtain for the potential and components
of the electric field in cylindrical coordinates the
quadrature expressions
\begin{equation}
\phi (\rho ,z) = {2d\over R^2} \int\limits_0^\pi
\left[ {1 - \tilde\rho \cos\varphi \over
(1 + \tilde\rho^2 + \tilde z^2 - 2\tilde\rho \tilde z \cos\varphi )^{3/2}}
\right] \, d\varphi \, .
\label{5-1-0}\end{equation}
\begin{equation}
E_{\rho}(\rho ,z) = {2d\over R^3} \int\limits_0^\pi
\left[ {\cos\varphi (\tilde z^2 - 2\tilde\rho^2 -2) + 3\tilde\rho
+\tilde\rho \cos^2\varphi \over
(1 + \tilde\rho^2 + \tilde z^2 - 2\tilde\rho \tilde z \cos\varphi )^{5/2}}
\right] \, d\varphi \, .
\label{5-1-3}\end{equation}
\begin{equation}
E_z(\rho ,z) = {2d\over R^3} \int\limits_0^\pi
\left[ { 3\tilde z (1 - \tilde\rho \cos\varphi ) \over
(1 + \tilde\rho^2 + \tilde z^2 - 2\tilde\rho \tilde z \cos\varphi )^{5/2}}
\right] \, d\varphi \, .
\label{5-1-4}\end{equation}

\begin{figure}[t]
\centering \epsfbox[10 500 200 700]{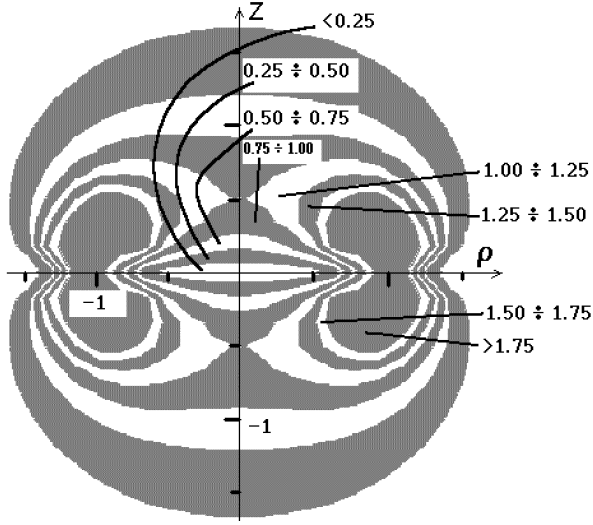} \label{r3}
\caption{Magnitude of the poloidal component of the electric field
$E_{||}(\rho /R, z/R)$ in units of its value at the saddle point.
Z and $\rho$ are expressed in fractions of $R$.}
\end{figure}

Here,  $\tilde\rho = \rho /R$ and  $\tilde z = z/R$ are
dimensionless cylindrical coordinates. The last three formulas are
expressed in terms of derivatives of the full elliptical
integrals. The result of this integration is shown in Fig. 2.

Acceleration by the electric field is associated only
with the component parallel to the magnetic field.
This acceleration is especially efficient on the $\Omega$ axis,
where the electric and magnetic lines of force are
parallel. We can readily see from (15) that the maximum
electric field on the $\Omega$ axis is reached at the
point $z = R/2$. However, this point is not a local
maximum of the modulus of the longitudinal
electric-field component
$E_{||} \equiv ({\bf EH})/H$.
Figure 3 shows the surface of the
$E_{||}$ force field, where we can see that
the point $z = R/2$ on the axis is a saddle point of the
distribution of $E_{||}$.
 With approach to the torus, the
conductivity of the medium should increase and the
field should become force-free: $E_{||}\to  0$.
\footnote{Note that, in a Blandford-Znajek model, the field is
forcefree everywhere by definition, so that efficient acceleration is
not possible.}

According to (13), the potential at the saddle point
is approximately $28 \%$
lower than the potential at
the center of the system, and is roughly an order of
magnitude lower than the maximum potential on the
torus surface. The extent of the region of acceleration
along the $z$ axis (at the level of $0.5$ of the value of
$E_{||}$ at the saddle point) is determined by the points
$z_1\approx 0.25,\,\,  z_2\approx 1.25$  in fractions of
$R$. Assuming that a mass $M$ with its corresponding

gravitational radius $r_g$ is at the center of the system,
we can use (13) to estimate the energy to which
particles with the elementary charge initially at the
saddle point on the axis can be accelerated:
\begin{equation}
E_k\approx \left({6.5\cdot a\over R}\right)
\cdot \left({\Omega R\over c}\right)\cdot \left({H_{_0}\over 10^4
\makebox{{\it Гс}}}\right)
\cdot \left({a\over r_g}\right)\cdot
\left({M\over 10^9 M_{^{_\bigodot}} }\right)
\cdot \left[\ln(4/b) - 2 \right] \cdot 10^{20} eV \, .
\label{5-2}\end{equation}
We can see that the factors in parantheses can be of
the order of unity in quasars, so that the kinetic energy
of particles accelerated by such a system can reach
$10^{20}$ eV.

\section{CONCLUSIONS}

We can draw the following conclusions from our
results.

1) The magnetic field at large distances approaches
a dipolar field, while the electric field corresponds
to a quadrupolar distribution for the charge
induced on the torus surface. It follows from Fig. 2
that, in the model considered here, in contrast to a
Blandford-Znajek model, we find a tendency for the
electric field lines to become more concentrated near
the $\Omega$ axis, which can explain the observed focusing
(collimation) of the accelerated relativistic particles.

2) The dimensions and locations of the regions of
cosmic-ray acceleration we have found can be used to
compare our results with observational data.

3) Thanks to its covariance, our method for the
computation of the electromagnetic field in a system
with toroidal symmetry can be generalized to the case
of a gravitationally curved space-time.

4) The mechanism considered here yields accelerated-
particle energies with the same order of magnitude
as the Blandford-Znajek mechanism (see, for
example, [1, 9, 10]).

\section{ACKNOWLEDGEMENTS}

This work was supported by the Russian Foundation
for Basic Research (project codes 01-02-16812,
00-15-96698, and 01-02-17829).

\section{APPENDICES}

$Finding\,\, the\,\, External\,\, Solution$

Let us write Maxwell's equations for arbitrary
curvilinear coordinates in the axially symmetric and
stationary case [7, $\S$90]:
\begin{equation}
\left\{
\begin{array}{lll}
e^{\alpha\beta\varphi}\partial_\beta F_{\alpha\varphi}=0\, ;\quad
\partial_\beta\left(\sqrt{-g}
g^{\alpha\beta}g^{\varphi\varphi}F_{\alpha\varphi}\right) =
4\pi\sqrt{-g}\, j^\varphi\, ;\\
e^{\alpha\beta\varphi}\partial_\beta F_{\alpha 0}=0\, ;\quad
\partial_\beta\left(\sqrt{-g}
g^{\alpha\beta}g^{00}F_{\alpha 0}\right) = 4\pi\sqrt{-g}\, j^0 \, .
\end{array}
\right.
\label{3-1}
\end{equation}
Here,  $e^{\alpha\beta\varphi}$  is a Levi-Civita symbol,
$\sqrt{-g}$ and $g^{ij}$ are
defined by expression (3), and
$j^i$  is the current 4-vector, which is identically equal to zero when
$x > a$.
We obtain for the magnetic field from (8) and the first
of equations (17):
\begin{equation}
F_{x\varphi}=-\partial_x R_n(x) \cos (\psi n)\, ,\quad
A_\varphi = - R_n(x) \cos (\psi n) .
\label{3-2}\end{equation}
Using (9), the external solution for the electric field
can be expanded in a Fourier series in the variable $\psi$.
Then, in accordance with the third equation of (17),
we obtain the solution for the electric field (for $x > a$)
\begin{equation}
\begin{array}{ccc}
F_{\psi 0}=nZ_n(x)\sin (\psi n) ,\quad
F_{x0}=-\partial_x Z_n(x) \cos (\psi n) ,\quad
A_0 = - Z_n(x) \cos (\psi n) .
\end{array}
\label{3-3}\end{equation}
It follows from (8) and (19) that the external electric
field is generated by the normal component of the
magnetic field at the torus boundary.
Everywhere where the expression
$(R - x\cos\psi)$
 is positive, such as in the range $a < x < R$, we can
remove the modulus signs in the second expression
of (3). With this in mind, introducing the dimensionless
variable $y = x/(2R)$ and denoting a derivative
with respect to this variable with a prime, we substitute
(19) into the fourth Maxwell equation (17) and
obtain
\begin{equation}
\begin{array}{ccc}
  y \left[ y^2 Z''_n +2y Z'_n -n(n-1) Z_n \right] \cos\{ (n-1)\psi\} -\\
- \left[ y^2 Z''_n +y Z'_n -n^2 Z_n \right] \cos\{\psi n\} +\\
+ y \left[ y^2 Z''_n +2y Z'_n -n(n+1) Z_n \right] \cos\{ (n+1)\psi\}
 = 0 \, .
\end{array}
\label{3-4}\end{equation}
The boundary conditions for
$Z_n(y)$
when $y = b \equiv a/(2R)$ and $n > 0$ follow from (9):
\begin{equation}
Z_n(b)=-\Omega R_n(b)\, , \quad ( n > 0 ) \, .
\label{3-3-1}\end{equation}
The necessary condition for the total electric charge
in the torus to be zero can be written
\begin{equation}
Q=\oint\limits F_{x0}\, g^{00}g^{xx}\,\sqrt{-g}\, 2\pi\, d\psi =0\quad
(\makebox{для }x\makebox{ в пределах: }\,\, a < x < R ) .
\label{3-4-1}\end{equation}
This expression flows from integration of the fourth
equation of (17). We thus obtain
\begin{equation}
Z'_0 = y Z'_1
\label{3-4-2}\end{equation}
Setting the expressions with the same harmonics
in (20) equal to zero, we obtain a recurrent system of
equations for $Z_n(y)$. Equation (23) is contained in the
Maxwell equation (20) with harmonic $n = 0$.
The condition that the total charge of the system
be equal to zero can also be written for $x > R$. For
this, we must take a contour passing around the
system, so that
$(R - x\cos\psi)$ does not change sign; i.e.,
the contour should be located to the right of the halfplane
from the $\Omega$ axis. We choose this contour as
shown in Fig. 4.
\begin{figure}[t]
\centering \epsfbox[100 500 300 700]{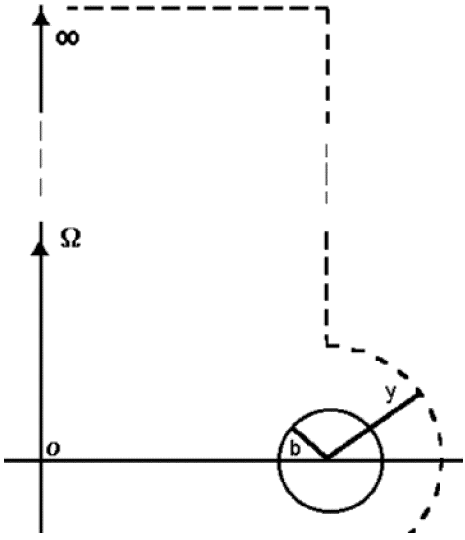} \label{r4}
\caption{Contour passing around the system, shown by the dashed
line.}
\end{figure}
 Thus, we pass by the torus along a
semicircle with radius $x > a$ on the right-hand side,
and we allow the contour to approach infinity at the
angles $\psi = \pi /2$ and $\psi = -\pi /2$
along half-lines parallel
and antiparallel to the $\Omega$ axis. The two (upper and
lower) "caps" of the finite area
$\pi R^2$ remain unclosed
at infinity, but since the field there asymptotically
approaches zero, the integrals on these "caps" can
be neglected. We can write the surface integral over
this contour in accordance with the fourth of equations
(17) taking into account the mirror symmetry
for the electric field in the equatorial plane:
\begin{equation}
Q=\int\limits_{\pi /2}^{\pi} F_{x0}\, g^{00}g^{xx}\,\sqrt{-g}\, 4\pi\, d\psi
+ \int\limits_{x}^{\infty} \left.
F_{\psi 0}\, g^{00}g^{\psi\psi}\, \sqrt{-g}
\, 4\pi\, dx \right|_{\psi =\pi /2} = 0 \, .
\label{3-4-3}\end{equation}
After straightforward but cumbersome manipulations,
we obtain
\begin{equation}
y Z'_n \left[ \pi (y\delta_n^1 - \delta_n^0 ) +
2y {\cos ({\pi n\over 2})\over n^2-1} +
{\sin ({\pi n\over 2})\over n} \right] + \sin\left( {\pi n\over 2}\right)
\int\limits_y^\infty {Z_n\over y}\, dy = 0\, .
\label{3-4-3-1}\end{equation}
Let us solve this problem with accuracy to within
the first three terms in the Fourier expansion
($Z_0,\, Z_1$, and $Z_2$).
In this case, taking into account (23), the
last expression becomes
\begin{equation}
(1-2y^2)y Z'_1 + \int\limits_y^\infty {Z_1\over y}\, dy \approx
{2\over 3} y^2 Z'_2 \, .
\label{3-4-4-0}\end{equation}
Differentiating, we obtain
\begin{equation}
(1-2y^2)y^2 Z''_1 + (1-6y^2)y Z'_1 - Z_1
\approx {2\over 3}y (y^2 Z''_2 + 2y Z'_2) \, .
\label{3-4-4}\end{equation}
We now write the Maxwell equation (20) corresponding
to the harmonic $n = 1$:
\begin{equation}
(1-2y^2)y^2 Z''_1 + (1-6y^2)y Z'_1 - Z_1
- y (y^2 Z''_2 + 2y Z'_2 - 2 Z_2) = 0 \, .
\label{3-4-4-1}\end{equation}
An equation for  $Z_2$ follows from these last two expressions:
\begin{equation}
y^2 Z''_2 + 2y Z'_2 - 6 Z_2 = 0 \, .
\label{3-4-5}\end{equation}
The solution vanishing at infinity has the form
\begin{equation}
Z_2 (y) = C_2 / y^3 \, .
\label{3-4-6}\end{equation}
The constant $C_2$ is found from the boundary conditions
(21). We can find $Z_1(y)$ using any of equations
(27) or (28):
\begin{equation}
(1-2y^2)y^2 Z''_1 + (1-6y^2)y Z'_1 - Z_1 = 4C_2 / y^2 \, .
\label{3-4-7}\end{equation}
A partial solution of the inhomogeneous equation (31)
has the form $Z_1^1 = 4C_2 /(3y^2)$.
 However, it does not
satisfy (26), of which (27) is a consequence, so that
we must search for a solution of $Z_1(y)$ from (26).
The substitution
$\int\limits_y^\infty (Z_1^o/y)\, dy = f_\infty - f(y)$
brings the homogeneous equation (26) into the form
$(1 - 2y^2) y (yf')' - f + f_\infty = 0$.
 Hence, in the limit $y \to 0$,
we obtain $Z_1^0 \to C_1 y^{\pm 1}$.\\
In the limit  $y\to\infty$,   the substitution
$y=\exp (\xi )$ reduces (26) to
$$
f''_о(\xi ) + e^{-2\xi}f_о /2=0\, ,\quad
f''_н(\xi ) + e^{-2\xi}f_н /2=e^{-2\xi}f_\infty /2 \, ,\quad
f=f_о + f_н \, .
$$
In the main approximation in $1/y$, the solution has the
form $f(y) = \exp [-1/(8y^2)] + 1/(8y^2)$.
 Hence, the asymptotic for  $y\to \infty$
has the form  $Z_1^0 \to -C_1/(32y^4)$.

$Z_0(y)$ can be found using (23) after $Z_1(y)$ is found.
We present the asymptotics of these solutions for
small and large $y$:
\begin{equation}
\begin{array}{ccc}
Z_0\mathop{\longrightarrow}\limits_{y\to 0} -
C_1\ln (y) + 8 C_2/(3y)\, ,\quad
Z_1\mathop{\longrightarrow}\limits_{y\to 0}
C_1 / y + 4 C_2/(3y^2)\, ,\\
Z_0\mathop{\longrightarrow}\limits_{y\to\infty} -
(C_1+8C_2)/(24y^3)\, ,\quad
Z_1\mathop{\longrightarrow}\limits_{y\to\infty} -
(C_1+8C_2)/(32y^4)\, .
\end{array}
\label{3-4-8}\end{equation}
Here, $C_1$ is the coefficient for the solution of the
homogeneous equation (26); like $C_2$, it is determined
by the specific configuration of the magnetic field from
the boundary conditions.
$$ $$

$Calculation\,\, of\,\, Coefficients\,\, for\,\, the\,\, Special\,\, Case$

$of\,\, a\,\, Current\,\, Ring\,\, in\,\, the\,\, Torus$

Let us now find the magnetic field in a simple case.
Let the current in the torus be in the form of a ring
along the torus axis with a delta-function distribution
[7, $\S$90]:
\begin{equation}
j^\varphi = \lim_{x_0\to 0} J^\varphi \delta (x-x_0)\, \delta (\psi -
\psi_0)
/ (2\pi\sqrt{-g})\, ,\quad j^\alpha = 0\, .
\label{4-1-0}\end{equation}
To find the potential $A_\varphi$ corresponding to this current,
we introduce the physical components of vectors in
accordance with the definition
$$
{\bf H_{phys}\equiv \hat H} = \{ H^\beta\sqrt{|g_{\alpha\beta}|}\} .
$$
This is necessary because the Biot-Savart law in its
usual form [7, $\S$43] is written in physical coordinates:
\begin{equation}
\hat A^\alpha {_{(x,\psi )}}=\hat e^\alpha
\displaystyle{\oint\limits_{}^{}}
{({\hat {\tilde j}}^\gamma\cdot\hat e_\gamma )\over |{\bf r- R}|}\,
\sqrt{-\tilde g}\, d\tilde x\, d\tilde\psi\, d\tilde\varphi\, .
\label{4-1}\end{equation}
Here, ${\bf e^\alpha}$ is a unit vector in the direction of the angle
$\varphi$, $(\hat{\tilde j^\gamma}\cdot \hat e_\gamma ) =
\tilde j^\varphi \sqrt{|g_{\varphi\varphi}|} \cos\tilde\varphi$,
and  $|{\bf r - R}|^2 = 2R (R - x\cos\tilde\varphi) + x^2$
is the square of the distance
from the segment of current with coordinate
$\tilde\varphi$  to the observation point.

Introducing the notation
$\tilde\varphi = \pi + 2\phi$,
$\kappa^2 = (1 - 2y \cos\psi )/(1 - 2y \cos\psi + y^2)$,
 we can transform  (34) to the form
$$
A_\varphi = (J^\varphi R/\pi)\sqrt{1-2y\cos\psi}\cdot \kappa \cdot
\int\limits_0^{\pi /2}{2\sin^2\phi -1\over\sqrt{1-\kappa^2\sin^2\phi}}\,
d\phi
$$
or
\begin{equation}
A_\varphi = 2(J^\varphi R/\pi)\cdot\sqrt{1-2y\cos\psi}\cdot
\left\{ K(\kappa )(1-\kappa^2/2)-E(\kappa )\right\} / \kappa\, .
\label{4-2-1}\end{equation}
Here,
$K(\kappa )$ and $E(\kappa )$
are full elliptical integrals.

We can find the asymptotic of (35) as $y \to 0$, or
equivalently as  $\kappa\to 1$.
Further, using (18), we can
obtain for the Fourier coefficients of the magnetic
field:
\begin{equation}
\begin{array}{ccc}
\lim\limits_{y\to 0}\, :
R_0 \to {J^\varphi R\over\pi}\left[2-\ln({4\over y}) \right]\, ,
\,
R_1 \to -{J^\varphi R\over\pi}\left[y\left\{ 1-\ln({4\over y})\right\}
\right]\, ,
\,
R_2 \to -{J^\varphi R\over\pi}\left[y^2
\left\{1-\ln({4\over y})/4 \right\}\right]\, .
\end{array}
\label{4-2-2}\end{equation}
It is clear from the boundary conditions (21) that,
according to (36),
$|Z_2(b)/Z_1(b)| \sim b \to 0$ as $b\to 0$,
so that we can neglect the remaining terms of
the Fourier series in the case of this magnetic-
field configuration. We introduce the notation
$H_0 \equiv |\partial_x \hat A^\varphi |_{\{ x=R,\, \psi = 0 \}} =
 J^\varphi /(2R)$.
for the magnetic-field
strength at the center of the system. We find the
coefficients in this approximation from
(21), (32), and (36):
\begin{equation}
C_1\approx 2\Omega H_{_0} R^2 b^2 \left[ 1- \ln(4/b) \right] /\pi \,
,\quad
C_2\approx 2\Omega H_{_0} R^2 b^5 \left[ 1 -\ln(4/b) / 4 \right] /\pi \,
.
\label{4-2-3}\end{equation}
$ $

\hrule

$$ $$
REFERENCES

[1]  R. D. Blandford and R. L. Znajek, Mon. Not. R.
Astron. Soc. 179, 433 (1977).

[2]  R. D. Blandford, astro-ph/0110396.

[3]  G. Barbiellini and F. Longo, astro-ph/0105464.

[4]  N. S. Kardashev, Mon. Not. R. Astron. Soc. 276, 515
(1995).

[5]  P. Goldreich andW. H. Julian, Astrophys. J. 157, 869
(1969).

[6]  J. Deutsch, Ann. Astrophys. 1 (1), 1 (1955).

[7]  L. D. Landau and E. M. Lifshitz, Course of Theoretical
Physics, Vol. 2: The Classical Theory of Fields
(Nauka, Moscow, 1988; Pergamon, Oxford, 1975).

[8]  L. D. Landau and E.M. Lifshitz, Course of Theoretical
Physics, Vol. 8: Electrodynamics of Continuous
Media (Nauka,Moscow, 1992; Pergamon, New York,
1984).

[9]  BlackHo les: theMembrane Paradigm, Ed. by K. S.
Thorne, R. H. Price, andD.A.Macdonald (YaleUniv.
Press, New Haven, 1986, Mir,Moscow, 1988).

[10]  A. A. Shatskiy,   Zh. Eksp. Teor. Fiz. 11, (2001).

$ $
\hrule
$ $

Translated by D. Gabuzda

ASTRONOMY REPORTS Vol. 46 No. 8 2002

\end{document}